\documentclass{article}
\usepackage[dvipdfmx]{graphicx,color}
\usepackage{amsmath,amssymb}

\begin{document}

\title{Simultaneous measurability of error and disturbance}
\author{Riuji Mochizuki\thanks{E-mail: rjmochi@tdc.ac.jp}\\
Laboratory of Physics, Tokyo Dental College,\\ Tokyo 101-0062, Japan }

\maketitle
\abstract{
The uncertainty relation, which displays an elementary property of quantum theory, was originally described by Heisenberg as the relation between error and disturbance.  Ozawa presented a more rigorous expression of the uncertainty relation, which was later  verified experimentally.  Nevertheless, the operators corresponding to error and  disturbance should be  measurable in the identical state if we follow the presupposition of Heisenberg's thought experiment.  In this letter, we discuss simultaneous measurability of error and disturbance and present a new inequality using error and disturbance in the identical state.  A testable example of this inequality is also suggested.

\newpage
\section{Introduction}

The uncertainty relation, which displays an elementary property of quantum theory, was originally  described by Heisenberg\cite{Heisenberg} as the relation between the error $\epsilon$ and disturbance $\eta$ of a particle's position and momentum  as

\begin{equation}
\epsilon \eta \ge h,\label{eq:Heisenberg}
\end{equation}
where $h$ is Planck's constant.

Subsequently, a more generalized inequality was shown\cite{Kennard}\cite{Robertson}: 
\begin{equation}
\sigma (A)\sigma (B)\ge {1\over 2} |\langle [A,B]\rangle |,\label{eq:Robertson}
\end{equation}
where $\sigma (X)$ is
the standard deviation of a self-conjugate operator $X$, which corresponds to some physical quantity, defined as 

\begin{equation}
\sigma (X) = \langle(\Delta X)^2\rangle^{1/2},\label{eq:sigma}
\end{equation}
with
\begin{equation}
\Delta X = X^{in} - \langle X^{in}\rangle ,\label{eq:Delta}
\end{equation}
and $[A, B]$ as the commutator of $A$ and $B$.

In some literature (for example, \cite{Neumann}), (\ref{eq:Robertson}) is considered to be a more formal expression of (\ref{eq:Heisenberg}).

Several decades later, Ozawa presented a more rigorous expression of the uncertainty relation\cite{Ozawa1}\cite{Ozawa2}\cite{Ozawa2.5}. The root-mean-square noise $\epsilon (A)$ and root-mean-square disturbance $\eta (B)$ are defined as 
\begin{equation}
\epsilon (A) = \langle N(A)^2\rangle^{1/2},\label{eq:epsilon}
\end{equation}
\begin{equation}
\eta (B) = \langle D(B)^2\rangle^{1/2}.\label{eq:eta}
\end{equation}
The Noise operator $N(A)$ is defined using the meter-observable $M^{out}$ of $A^{in}$ as
\begin{equation}
N(A)=M^{out}-A^{in},     \label{eq:N}
\end{equation}
with the disturbance operator $D(B)$ as 
\begin{equation}
D(B)=B^{out}-B^{in}, \label{eq:D}
\end{equation}
where $in$ and $out$ mean $just\ before$ and $just\ after$ measurement, respectively.  The new uncertainty relation is written by means of (\ref{eq:epsilon}), (\ref{eq:eta}) and also (\ref{eq:sigma}) as

\begin{equation}
\epsilon(A)\eta(B)+\epsilon(A)\sigma(B)+\sigma(A)\eta(B)\ge{1\over 2}|\langle [A^{in},B^{in}]\rangle | .\label{eq:Ozawa}
\end{equation}

Recently, it was reported\cite{Erhart} that (\ref{eq:Ozawa}) was verified experimentally by a neutron spin experiment.  Nevertheless, it is not clear whether verification of (\ref{eq:Ozawa}) is possible for continuous quantities such as position and momentum.  In other words,  it  is not clear whether  (\ref{eq:epsilon}) and  (\ref{eq:eta}) are measurable for such quantities\cite{Werner}\cite{Shimizu}.  Watanabe et  al.\cite{Watanabe3}\cite{Watanabe1}\cite{Watanabe2} suggested another inequality suitable for practical measurement.

Moreover, error and disturbance were defined in the identical state in Heisenberg's thought experiment\cite{Heisenberg} referring to the uncertainty principle.  If we follow his presupposition, the operators  corresponding to error and disturbance should be simultaneously measurable.  In many textbooks on quantum theory, commutativity of observables is regarded as a necessary and sufficient condition of possibility of simultaneous measurement.  Ozawa, however, insists in his paper\cite{Ozawa3} that, in some states, two noncommutative observables, $A$ and $B$, are simultaneously measurable if they satisfy    
\begin{equation}
\epsilon(A)=\epsilon(B)=0 \label{eq:ee}
\end{equation}
and their meter observables are commutative.  Simultaneous measurability has been discussed with respect to contextuality and weak measurement\cite{Ozawa3}\cite{Dressel1}\cite{Dressel2}\cite{Ozawa4}.

The purpose of this letter is to discuss the simultaneous measurability of error and disturbance.  Firstly, we define simultaneous measurability from the quantum logical aspect.  According to our definition, there exists no state where noncommutative observables are simultaneously measurable.  Then, we define commutative operators which correspond to the error and disturbance of noncommutative observables.  This definition leads to the uncertainty relation of error and disturbance in the identical state.   A testable example of this relation is also suggested, where definition of error $\epsilon$ in \cite{Erhart} is shown to be insufficient for other settings.

\section{Simultaneous measurability}

To prepare for discussion about simultaneous measurability, we define {\it observables} according to a common quantum logical approach\cite{Svozil}\cite{Maeda}.  The proposition that a measured value of a physical quantity $u$ belongs to a subspace $A$ of space of real number ${\bf R}$ is written as  $u(A)$.  When the truth value of $u(A)$ can be determined experimentally, $u$ is called measurable.  Logic $L$, which is nothing but a $\sigma$-complete orthomodular lattice, consists of such propositions.   Classical logic is a  Boolean lattice, namely, an orthocomplemented distributive lattice, while quantum logic is not.

We suppose $\sigma$-field $\mathcal{B}$({\bf R}), which consists of all open sets belonging to space of real number {\bf R}.  A map $u$ from $\mathcal{B}$({\bf R}) to  logic $L$ is called an observable of $L$ if
\begin{equation}
u({\bf R})=1,\ \ u(\emptyset)=0,\label{eq:1}
\end{equation}
\begin{equation}
u(A)^{\bot}=u({\bf R}-A)\ {\rm for}\ A\in \mathcal{B}({\bf R}),
\end{equation}
\begin{equation}
u(\bigcup^{\infty}_{n=1}A_n)=\bigvee^{\infty}_{n=1}u(A_n)\ {\rm for}\ A_n\in \mathcal{B}({\bf R}), \ {\rm  if}\  A_m\cap A_n=\emptyset \ {\rm for}\ m\neq n,\label{eq:3}
\end{equation}

where $u(A)^{\bot}$ is the orthocomplement of $u(A)$ and  $\{u(A_n)\ ;\ n=1,2,\cdots\}$ constitute an orthogonal set of projection operators.  It is proved that observables are $\sigma$-homomorphism from $\mathcal{B}({\bf R})$ to $L$.

There exists a one-to-one correspondence between the whole set of bounded observables and the whole set of bounded self-conjugate linear operators. 
If, and only if, two such operators, which correspond to observables $u$ and $v$, are commutative, they satisfy for any pair of $A,B\in \mathcal{B}({\bf R})$
\begin{equation}
v(B)=(v(B)\wedge u(A))\vee (v(B)\wedge (u(A))^{\bot})\label{eq:5}
\end{equation}
and the orthomodular lattice whose elements are $u(A)$'s and $v(B)$'s is Boolean. Here, we assume,  as usual, that all the measurable quantities are observables.

We define the simultaneous measurability of observables $u$ and $v$ as follows. \\
{\it Definition} 
 
$u$ and $v$ are called simultaneously measurable if the truth value of $u(A)\wedge v(B)$ can be determined experimentally.  

We present the following theorem:\\
{\it Theorem}

Let $u$ and $v$ be observables of logic $L$ and 
$u_{(v=B)}(A_n)\equiv u(A_n)\wedge v(B)\in L,\ A_n,B\in \mathcal{B}({\bf R}),\ \ n=1,2,\cdots$
for the fixed $v(B)$.   Then, $u_{(v=B)}(A_n),\ n=1,2,\cdots$ are observables if, and only if, they satisfy  (\ref{eq:5}).    \\  
{\it Proof} (sufficiency)

We assume  (\ref{eq:5}) is satisfied.  Firstly, we show the whole set $L_{v=B}$ whose elements are  $u_{(v=B)}(A_n),\ n=1,2,\cdots$ is a $\sigma$-complete orthocomplemented distributive lattice.  Since $u(A_n)$'s and $v(B)$ satisfy the distribution law,   
\[
\bigvee_n u_{(v=B)}(A_n)=\big(\bigvee_n u(A_n)\big)\wedge v(B)\in L_{v=B}
\]
and $u_{(v=B)}(A_n)$ also satisfy the distribution law.  Moreover, if we define 
\begin{equation}
\big(u(A)\wedge v(B)\big)^{\bot} \equiv \big(u(A)\big)^{\bot}\wedge v(B)          
\end{equation}
for $u(A)\wedge v(B)\in L_{v=B}$, $\big(u(A)\wedge v(B)\big)^{\bot}$ is the  orthocomplement of $u(A)\wedge v(B)$.  Thus $L_{v=B}$ is a $\sigma$-complete orthocomplemented distributive lattice.  It is clear that $u_{(v=B)}(A_n),\ n=1,2,\cdots$ satisfy (\ref{eq:1})$\sim$(\ref{eq:3}) because $L_{v=B}$ is a distributive lattice.  Therefore $u_{(v=B)}(A_n),\ n=1,2,\cdots$ are observables of $L_{v=b}$ if they satisfy (\ref{eq:5}). \\
(necessity)

Let $u_{(v=B)}(A_n),\ n=1,2,\cdots$ be observables.  From (\ref{eq:3})
\[
u_{(v=B)}(A_m)\vee u_{(v=B)}(A_n)=u_{(v=B)}(A_m\cup A_n),
\]
if $A_m\cap A_n =\emptyset$.
This equation leads to
\[
\big( v(B)\wedge u(A_m)\big) \vee \big( v(B)\wedge u(A_n)\big) =
v(B)\wedge u(A_m\cup A_n)=v(B)\wedge \big(u(A_m)\vee u(A_n)\big).
\]
If we put $A_n ={\bf R}-A_m$,
\[
\big( v(B)\wedge u(A_m)\big) \vee \big( v(B)\wedge u({\bf R}-A_m)\big) =v(B)\wedge \big(u(A_m)\vee u(A_m)^{\bot}\big)=v(B).
\]
QED.

From the above,  it is shown that $u(A)\wedge v(B)$ is not an observable if  (\ref{eq:5}) is not satisfied, that is, two observables which correspond to mutually-noncommutative linear operators are not simultaneously measurable.

For example, let 
\[
P_{x+}={1-\sigma_x\over 2},
\]
\[
P_{\phi +}={1-\sigma_{\phi}\over 2},
\]
be projection operators corresponding to $u(A)$ and $v(B)$, respectively, where 
\[
\sigma_{\phi}=\sigma_x\cos\phi + \sigma_y\sin\phi.
\]
 $\sigma_x$ and $\sigma_y$ are Pauli spin matrices.  Then, if $\phi\neq 0$, the projection operator corresponding to $u(A)\wedge v(B)$ is $0$, which is not an observable. 

\section{Uncertainty relation}

From the previous section, we can say such quantities as
\begin{equation}
\langle N(A) D(B)\rangle,\label{eq:ND}
\end{equation}
are not measurable because (\ref{eq:N}) and  (\ref{eq:D}) are noncommutative when $[A,B]\ne 0$.  Note that this fact does not deny (\ref{eq:Ozawa}) where (\ref{eq:ND}) does not appear but (\ref{eq:epsilon}), (\ref{eq:eta}) and (\ref{eq:sigma}) do.  These are measured separately by using states belonging to the same statistical ensemble.  What we would like to emphasize is that the uncertainty relation should be written by means of commutative quantities if it is thought to be the relation between quantities which are measured in the identical state.  Thus we define 
\begin{equation}
{\cal N}(A)=M^{out} - \langle A^{in}\rangle,
\end{equation}
\begin{equation}
{\cal D}(B) = B^{out} - \langle B^{in}\rangle ,
\end{equation}
as operators which express error and disturbance from the expectation values, respectively.

Using these operators, we examine the following quantity:
\begin{equation}
\langle {\cal N}(A)^2{\cal D}(B)^2\rangle ^{1/2}.\label{eq:ND2}
\end{equation}
Since $M^{out}$ and $B^{out}$ are observables in different systems,  (\ref{eq:ND2}) becomes
\[
\langle {\cal N}(A)^2{\cal D}(B)^2\rangle ^{1/2}=\langle {\cal N}(A)^2\rangle^{1/2}\langle{\cal D}(B)^2\rangle ^{1/2}.
\]
If we use 

\begin{equation}
\langle {\cal N}(A)^2\rangle ^{1/2}=\langle (N(A)+\Delta A)^2\rangle^{1/2},
\end{equation}
\begin{equation}
\langle{\cal D}(B)^2\rangle ^{1/2}=\langle (D(B)+\Delta B )^2\rangle^{1/2}.
\end{equation}
and assume 
\begin{equation}
\langle N(A)\Delta A\rangle = \langle D(B)\Delta B\rangle = 0,\label{eq:assumption}
\end{equation}
(\ref{eq:ND2}) is written by the use of (\ref{eq:sigma}), (\ref{eq:epsilon}) and (\ref{eq:eta}) as
\begin{equation}
\langle {\cal N}(A)^2{\cal D}(B)^2\rangle ^{1/2}=(\epsilon (A)^2 + \sigma (A)^2)^{1/2}(\eta (B)^2+\sigma (B)^2)^{1/2}.\label{eq:ND3}
\end{equation}
It is clear that (\ref{eq:assumption}) is not invariably realized.  One of the simplest counter examples is the case where $M^{out}$ always indicates $\langle A^{in}\rangle $.  Nevertheless, we regard (\ref{eq:assumption}) as  a rather reasonable assumption, which means that $N(A)$ and $\Delta A$ are independent stochastic variables, and so are $D(B)$ and $\Delta B$,  


We can calculate the lower bound of (\ref{eq:ND3}) by means of (\ref{eq:Robertson}) and (\ref{eq:Ozawa}) to obtain

\begin{equation}
\langle {\cal N}(A)^2{\cal D}(B)^2\rangle ^{1/2} \ge (2-\sqrt{2})|\langle [A,B]\rangle |.\label{eq:Mochi}
\end{equation}
If we use

\begin{equation}
\epsilon (A)\eta (B)\ge {1\over 2} |\langle [A,B]\rangle | \label{eq:Heis}
\end{equation}
in place of (\ref{eq:Ozawa}), the minimal value becomes almost double:

\begin{equation}
\langle {\cal N}(A)^2{\cal D}(B)^2\rangle ^{1/2} \ge |\langle [A,B]\rangle |.\label{eq:Mochi2}
\end{equation}

\section{A testable example}

In this section, we suggest an experiment with a setting which is a little modified from the experiment in \cite{Erhart} as a testable example of the inequality (\ref{eq:Mochi}).  We define $A$, $B$ and $O_A$ instead of their definition in \cite{Erhart} as
\begin{equation}
A=O_A=\sigma_x \sin\theta + \sigma_z\cos\theta,
\end{equation}
\begin{equation}
B={\sigma_y\over\sqrt{2}}+{\sigma_z\over\sqrt{2}},
\end{equation}
where
\[
0\leq\theta\leq{\pi\over 2}
\]
and $\psi\rangle =|+z\rangle$.  (\ref{eq:assumption}) , which is necessary to conclude with (\ref{eq:Mochi}), is satisfied in this setting.  If the root-mean-square noise $\epsilon(A)$ is completely calculable by using $A$, $B$ and $O_A$ as insisted in \cite{Erhart}, 
\begin{equation}
\sigma (A)= \sin \theta,
\end{equation}
\begin{equation}
\sigma (B) = {1\over \sqrt{2}},
\end{equation}
\begin{equation}
\epsilon (A)=0,
\end{equation}
\begin{equation}
\eta (B) = \sin \theta.
\end{equation}
Then, 
\begin{equation}
\epsilon(A)\eta(B)+\epsilon(A)\sigma(B)+\sigma(A)\eta(B)=\sin ^2\theta
\end{equation}
and
\begin{equation}
|\langle [A^{in}, B^{in}]\rangle |=\sqrt{2}\sin\theta.
\end{equation}
It comes down to that Ozawa's inequality (\ref{eq:Ozawa}) is not realized within $\sin\theta<1/\sqrt{2}$.  This fact seems to show that $\epsilon(A)$ includes uncontrollable error.

Accordingly, we will estimate the range of $\epsilon (A)$, including uncontrollable error, on the assumption that (\ref{eq:Heis}) or (\ref{eq:Ozawa}) is realized.  We redefine $\epsilon (A)$ as 
\begin{equation}
\epsilon (A) = \langle (M^{out}+\delta M-A^{in})\rangle^{1/2},
\end{equation}
where $\delta M$ is the operator which gives uncontrollable error and is assumed to satisfy
\[
\langle \delta M (A^{in}-\langle A^{in}\rangle\rangle=0.
\]
This assumption may demand that the angular momentum of the particle should be measured continuously.  Then, inequalities corresponding to (\ref{eq:Mochi2}) and (\ref{eq:Mochi}) will be derived from (\ref{eq:ND3}).

Firstly, if we assume (\ref{eq:Heis}), $\epsilon (A)\ge {1\over\sqrt{2}}$
 independently of $\theta$.  Then,
\begin{equation}
\langle {\cal N}(A)^2{\cal D}(B)^2\rangle ^{1/2} \ge \Big( {1\over 2\sqrt{2}\sin\theta}+{\sin\theta\over\sqrt{2}}\Big) |\langle [A,B]\rangle |\label{eq:Mochi21}
\end{equation}
The minimum value of the coefficient of the right-hand side is 1 when $\sin \theta = 1/\sqrt{2}$. 

Next,  if (\ref{eq:Ozawa}) is assumed, 

\[
\epsilon (A)  \ge 0, \ \ \ \ \ \ (\sin\theta\ge{1\over\sqrt{2}}).
\]
\begin{equation}
\epsilon (A)  \ge{\sin\theta(1-\sqrt{2}\sin\theta)\over 1+\sqrt{2}\sin\theta},\ \ \ \   \ \ (\sin\theta\le{1\over\sqrt{2}}).
\end{equation}
 Then
\[
\langle {\cal N}(A)^2{\cal D}(B)^2\rangle ^{1/2} \ge{1\over\sqrt{2}} (\sin^2\theta +{1\over 2})^{1/2} |\langle [A,B]\rangle |,\ \ \ \ \ \ (\sin\theta\ge{1\over\sqrt{2}}).\ \ \ \ \ 
\]
\begin{equation}
\langle {\cal N}(A)^2{\cal D}(B)^2\rangle ^{1/2} \ge\Big({1+2\sin^2\theta\over \sqrt{2}+2\sin\theta}\Big) |\langle [A,B]\rangle |,\ \ \ \ \ \ (\sin\theta\le{1\over\sqrt{2}}). \label{eq:Mochi11}
\end{equation}
The minimum value of the coefficient of the right-hand side is $2-\sqrt{2}$ when $\sin \theta = 1-1/\sqrt{2}$. 

If 
\begin{equation}
{\langle {\cal N}(A)^2{\cal D}(B)^2\rangle ^{1/2} \over |\langle [A,B]\rangle |}\le 1
\end{equation}
at some angles and 
\begin{equation}
2-\sqrt{2}\le{\langle {\cal N}(A)^2{\cal D}(B)^2\rangle ^{1/2} \over |\langle [A,B]\rangle |}
\end{equation}
at each angle are shown experimentally, we can conclude that the inequality (\ref{eq:Mochi}) is realized.   This is also an experimental proof that Ozawa's inequality is correct.



\end{document}